\documentclass[aps,pra,pdflatex,reprint,10pt,showpacs,superscriptaddress]{revtex4-2}
\usepackage{amssymb,bbold}
\usepackage{amsmath}
\usepackage[utf8x]{inputenc}
\usepackage{natbib}
\usepackage{graphicx}
\usepackage{hyperref}
\usepackage{epstopdf}
\usepackage{pifont}
\usepackage{float}
\usepackage{ulem}
\usepackage{booktabs}
\usepackage{soul}
\usepackage{orcidlink}

\graphicspath{{pdf/}}
\hypersetup{colorlinks=true,linkcolor=blue,citecolor=blue,filecolor=blue,urlcolor=blue,breaklinks=true}
\RequirePackage{color}

\begin{document}

\date{\today}

\title{Geometry-induced Coulomb-like potential in helically twisted quantum systems}

\author{Frankbelson dos S. Azevedo\orcidlink{0000-0002-4009-0720}}
\email[Frankbelson dos S. Azevedo - ]{frfisico@gmail.com}
\altaffiliation{On leaving the current affiliation and currently unaffiliated.}
\affiliation{Departamento de F\'{\i}sica, Universidade Federal do Maranh\~{a}o, 65085-580 S\~{a}o Lu\'{\i}s, Maranh\~{a}o, Brazil}

\author{Faizuddin Ahmed\orcidlink{0000-0003-2196-9622}}
\email[Faizuddin Ahmed - ]{faizuddinahmed15@gmail.com}
\affiliation{Department of Physics, Royal Global University, Guwahati, 781035, Assam, India}

\author{Edilberto O. Silva\orcidlink{0000-0002-0297-5747}}
\email[Edilberto O. Silva - ]{edilberto.silva@ufma.br}
\affiliation{Departamento de F\'{\i}sica, Universidade Federal do Maranh\~{a}o, 65085-580 S\~{a}o Lu\'{\i}s, Maranh\~{a}o, Brazil}

\begin{abstract}
In this paper, we investigate the Schrödinger equation in a three-dimensional helically twisted space characterized by a non-trivial torsion parameter. By applying exact separation of variables, we derive the radial equation governing the dynamics of quantum particles in this geometric background. Remarkably, the intrinsic coupling between angular and longitudinal momenta induced by the torsion gives rise to an attractive Coulomb-like potential term that emerges purely from the underlying geometry, without introducing any external fields or interactions. We obtain exact analytical solutions for the bound states, including both the energy spectrum and the normalized wave functions. Numerical calculations are also performed, and excellent agreement with the exact results is found. The interplay between the torsion parameter and the effective Coulomb-like interaction is analyzed in detail, showing how geometric deformation generates novel quantum confinement mechanisms in twisted spaces.\\

{\bf keywords:} 03.65.Nk--non-relativistic quantum system; 04.62.+v--Quantum fields in curved space; 02.30.Jr --partial differential equation; 03.65.Ge--solutions of wave equations: bound states

\end{abstract}

\maketitle

\section{Introduction}

Quantum mechanics in curved geometries constitutes a vibrant field of research with diverse applications. The interplay between quantum mechanics and curved geometries has revealed profound connections between spatial structure and quantum phenomena \cite{audretsch2012quantum}. The geometrization of quantum mechanics postulate that particle evolution follows geodesics in a curved space induced by the quantum potential \cite{TAVERNELLI2016239}. Quantum wave equations formulated for particles in curved spacetime have been applied to describe the motion of non-relativistic charged particles in the spacetime of a global monopole \cite{ALVES2024609}, the hydrogen atom spectrum in radial orbits of the Schwarzschild metric \cite{PhysRevD.25.3180}, and the two-dimensional hydrogen atom spectrum in the presence of a cosmic string \cite{DOSSAZEVEDO2024169660}.

While much attention focuses on conventional Riemannian manifolds, spaces with intrinsic torsion offer richer mathematical structures and novel physical effects under active exploration. Nontrivial geometries can induce effective interactions on quantum particles, even without external fields, resulting in geometry-induced potentials absent in flat space \cite{Venzo1991,Kleinert1989}. Research indicates that torsion generates additional geometric potentials in quantum strip waveguides \cite{IainJClark_1996} and for particles constrained to space curves \cite{PhysRevB.91.245412}. Torsion also influences spin-orbit coupling, introducing additional terms dependent on both torsion and curvature \cite{PhysRevA.97.042108}. In addition, geometric torsion in a twisted quantum ring can induce a persistent current \cite{Taira_2010}, and twisted graphene nanoribbons yield a refractive index controlled by the number of twists \cite{GurtasDogan_2025}. Finally, recent studies conducted by two of the present authors have investigated the Klein–Gordon equation and geodesics in the spacetime of a wiggly cosmic string with dislocation and the result is radial confinement \cite{dos2024dynamics}.

From a physical perspective, curved spacetime with torsion finds applications in distinct areas of physics. In condensed matter, it can model crystals containing screw dislocations, where the torsion parameter is directly related to the Burgers vector density characterizing the defect \cite{moraes2000condensed,PhysRevE.109.L012701}. Such geometry also appears in curved nanotubes and curved graphene \cite{de2021geometric}. In gravitational physics, this geometry appears in exact solutions with torsional stresses, serving as effective models for cosmic string-like configurations \cite{PhysRevD.53.779,jusufi2016light}. In emerging quantum materials, effective two-dimensional reductions of this metric may describe twisted bilayer systems, where interlayer coupling generates moiré structures with topological richness \cite{cao2018correlated}.

Non-relativistic quantum mechanics have been studied in three-dimensional curved spaces in the presence of various types of topological defects. Notable studies include the analysis of the harmonic oscillator in an elastic medium with spiral dislocation \cite{Maia2018}, the impact of spiral dislocation topology on the revival time of the harmonic oscillator \cite{Maia2022}, and the harmonic oscillator in a space-time where a vertical line is distorted into a vertical spiral \cite{Silva2019}. Other significant works explore the effects of screw dislocation and linear potentials on the harmonic oscillator \cite{Bueno2016}, the interaction of the harmonic oscillator with conical singularities \cite{Furtado2000}, and the behavior of the harmonic oscillator in spaces with linear topological defects \cite{Azevedo2001} or point-like defects \cite{Vitória_2019}. Additionally, studies address non-inertial effects on a non-relativistic quantum harmonic oscillator in the presence of screw dislocations \cite{Santos2023}, as well as the quantum dynamics of non-relativistic particles in wormhole backgrounds featuring disclinations \cite{Ahmed2023a, Ahmed2023b}. Other noteworthy studies include the investigation of rotational effects on electrons (or holes) confined in a quantum ring potential in the presence of a screw dislocation \cite{Dantas2015}, as well as the influence of screw dislocations on the energy levels of electrons in three-dimensional solid-state systems \cite{Cleverson2016}, quantum particles interacting with linear plus Coulomb potential and Kratzer potential under the Generalized Uncertainty Principle (GUP) considering the cosmic string effects \cite{Wang2016}, rotating effects on the Landau quantization for an atom with a magnetic quadrupole moment \cite{Fonseca2016}, interaction of an electron with a nonuniform axial magnetic field in a uniformly rotating frame \cite{Bakke2019}, interaction of a point charge with a uniform magnetic field in the distortion of a vertical line into a vertical spiral spacetime \cite{Silva-Bakke2019}. Collectively, these investigations provide valuable insights into the interplay between quantum mechanics and topological defects, deepening our understanding of quantum system behavior in complex and non-trivial space-time geometries. Exploring such effects promises to enhance our understanding of quantum phenomena across various physical contexts.

In this work, we investigate a paradigmatic case of such geometry: a stationary spacetime with helical torsion that emerges naturally. Remarkably, the helical structure induces a Coulomb-like potential term purely from the geometry, without introducing any external interaction. This geometry provides a unique framework for analyzing how quantum confinement and energy quantization can emerge solely from topological and geometrical deformations. This purely geometric mechanism leads to the formation of bound states, modifying both the energy spectrum and the spatial distribution of the wave functions. The emergence of a $(1/r)$-type potential from torsion-induced coupling between angular and longitudinal momenta represents a novel feature with no counterpart in flat space quantum mechanics.

The paper is organized as follows: in Sec.~\ref{model}, we present the geometric model and show the explicit form of the helical metric that incorporates torsion. In Sec.~\ref{equation}, we formulate the corresponding Schrödinger equation, perform the separation of variables, and derive the radial equation containing the geometry-induced effective potential. The exact analytical solutions for both the energy eigenvalues and normalized wave functions are obtained in Sec.~\ref{solutions}. In Sec.~\ref{numerical}, we present a detailed numerical analysis comparing the exact and numerical solutions for both energy levels and probability densities, confirming the accuracy of the analytical results. Finally, Sec.~\ref{conclusion} summarizes the main conclusions and discusses possible extensions of the model.

\section{Geometric Model: Helically Twisted Space \label{model}}

We consider a quantum system embedded in a nontrivial geometry characterized by helical torsion. The spatial part of the line element is written as \cite{Silva2025Helically}
\begin{equation}
ds^2 = dr^2 + r^2\, d\phi^2 + (dz + \omega\, r\, d\phi)^2, \label{metric}
\end{equation}
where $\omega$ is a constant parameter controlling the strength of the torsion, which is dimensionless. The cross-term $(dz + \omega\, r\, d\phi)^2$ reflects the screw-like structure of the space, coupling the angular coordinate $\phi$ to the longitudinal coordinate $z$. This geometry introduces an effective interaction purely of geometric origin, even in the absence of external fields.
Expanding the metric (\ref{metric}), we get
\begin{align}
ds^2 = dr^2 + r^2\,(1 + \omega^2)\, d\phi^2 + 2\, \omega\, r\, dz\, d\phi + dz^2.
\end{align}
From this expression, the spatial metric tensor $g_{ij}$, written in coordinates $(r, \phi, z)$, takes the matrix form
\begin{equation}
g_{ij} =
\begin{pmatrix}
1 & 0 & 0 \\
0 & r^2\,(1+\omega^2) & \omega\, r \\
0 & \omega\, r & 1
\end{pmatrix}.
\end{equation}
The determinant of this metric is readily calculated as $g = \det(g_{ij}) = r^2$, and the corresponding volume element becomes $\sqrt{g} = r$. The inverse metric components $g^{ij}$ are obtained as
\begin{equation}
g^{ij} =
\begin{pmatrix}
1 & 0 & 0 \\
0 & \dfrac{1}{r^2} & -\dfrac{\omega}{r} \\
0 & -\dfrac{\omega}{r} & 1+\omega^2
\end{pmatrix}.
\end{equation}

For more about the geometric foundations and energy conditions of the spacetime described above, see Ref. \cite{Silva2025Helically}. In what follows, we interpret the effects of such a spacetime on non-relativistic quantum particles.

\section{Schrödinger Equation and Emergence of the Coulomb-like Potential \label{equation}}

The time-independent Schrödinger equation in curved space is written using the Laplace-Beltrami operator as \cite{LDL}
\begin{equation}
E\,\Psi ({\bf r})= -\frac{\hbar^2}{2\mu}\,\frac{1}{\sqrt{g}}\, \partial_i \left( \sqrt{g}\, g^{ij}\, \partial_j \Psi ({\bf r}) \right)\quad (i,j=1,2,3), \label{schrodinger}
\end{equation}
where $\mathrm{E}$ is the particles energy, $\hbar$ is the planck's constant, $\mu$ is the effective mass of the particle, $g_{ij}$ is the spatial metric tensor with $g = \det(g_{ij})$ its determinant, and $\Psi$ is the spatial-dependent wave function. 

Employing the metric components obtained in the previous section, the Laplace-Beltrami operator takes the form
\begin{align}
E\,\Psi ({\bf r}) = -\frac{\hbar^2}{2\mu} \Bigg[ 
\frac{1}{r} \partial_r \left( r\, \partial_r\right) 
+ \frac{1}{r^2} \partial_\phi^2 
&- \frac{2\,\omega}{r} \partial_\phi\, \partial_z 
\notag \\ &+ (1+\omega^2) \partial_z^2\Bigg]\,\Psi ({\bf r}). \label{schrodinger_expanded}
\end{align}

Since the spatial metric tensor $ g_{ij} $ is independent of the coordinates $ \phi $ and $ z $, the wave function $ \Psi({\bf r}) $ can be expressed in terms of variables that allow for separation of variables. Accordingly, we adopt a separable solution of the form: 
\begin{equation}
\Psi(r,\phi,z) = e^{i\,m\,\phi}\, e^{i\,k\,z}\, \psi(r),\label{ansatz}
\end{equation}
where $m \in \mathbb{Z}$ is the azimuthal (orbital) quantum number associated with the angular coordinate and $k \in \mathbb{R}$ is the longitudinal momentum along the $z$-axis, and $\psi(r)$ is the radial function. Substituting this ansatz into Eq.~(\ref{schrodinger_expanded}), we obtain the radial equation
\begin{align}
E\,\psi(r) = -\frac{\hbar^2}{2\,\mu} \Bigg[ 
\frac{1}{r}\,\frac{d}{dr}\left( r\,\frac{d}{dr} \right)
&- \frac{m^2}{r^2}+\frac{2\,\omega\, k\,m}{r}
\notag \\&- (1+\omega^2)\, k^2\Bigg]\, \psi(r). \label{radial_equation_expanded}
\end{align}
This radial equation shows the essential physical structure generated by the helically twisted geometry. The first term represents the usual centrifugal barrier, while the second term introduces a Coulomb-like interaction proportional to $1/r$, which arises here exclusively due to the geometric coupling between the longitudinal and angular coordinates. This $1/r$ interaction is purely geometrical in nature and does not require the presence of any external potential. The last term also accounts for an energy renormalization associated with the torsion parameter $\omega$ and the longitudinal momentum $k$. 

After applying the transformation $ \psi(r) = r^{-1/2}\, f(r) $, which eliminates the first derivative, the radial equation takes the form
\begin{equation}
-\frac{\hbar^2}{2\mu} f''(r) + \mathcal{V}_{\mathrm{eff}}(r) f(r) = E f(r),
\end{equation}
where the effective potential $ \mathcal{V}_{\mathrm{eff}}(r) $ is given by
\begin{equation}
\mathcal{V}_{\mathrm{eff}}(r) = -\frac{\hbar^2}{2\mu} \left( - \frac{m^2-\frac{1}{4} }{r^2} + \frac{2\,\omega\, k\, m}{r} - (1+\omega^2)\,k^2 \right).\label{effective_potencial}
\end{equation}
This effective potential reveals two essential contributions: a modified centrifugal barrier, shifted by $-1/4r^2$, and a Coulomb-like attractive term proportional to $1/r$. Remarkably, this Coulomb-like interaction emerges solely from the torsional geometry, without requiring any external potential or interaction. The term proportional to $(1+\omega^2)k^2$ shifts the energy eigenvalues and plays no role in the effective potential structure.
\begin{figure}[tbhp]
\centering
\includegraphics[scale=0.45]{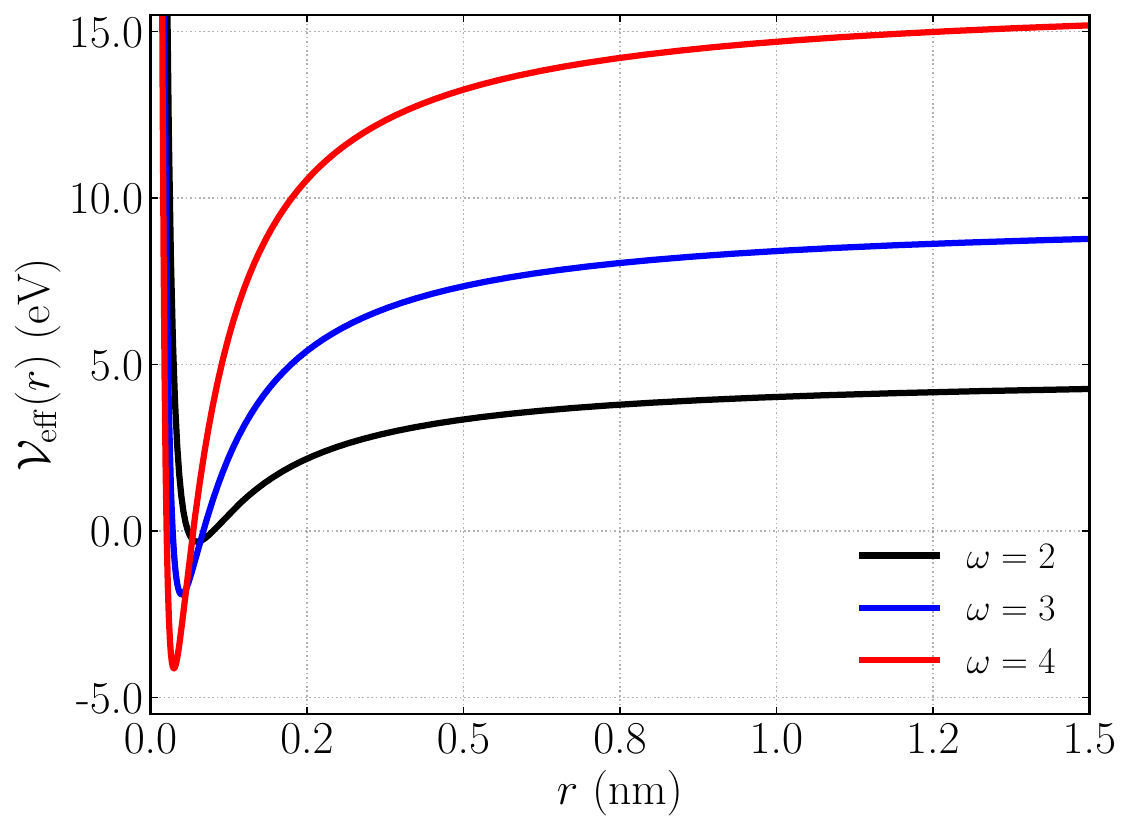}
\caption{\footnotesize Effective potential $\mathcal{V}_{\mathrm{eff}}(r)$ (Eq. (\ref{effective_potencial})) as a function of the radial distance $r$ for different values of the torsion parameter $\omega = 0.3$, $0.5$, and $0.7$. The potential exhibits a well-defined attractive region, whose depth increases with increasing $\omega$. This behavior directly reflects the enhancement of the $1/r$-type interaction induced by the torsional coupling between angular and longitudinal momenta. We use $m=1$ and $k_{z}=5.0\times10^{9}$ m$^{-1}$.}
\label{fig:Veff}
\end{figure}

Figure~\ref{fig:Veff} illustrates the behavior of the effective potential $\mathcal{V}_{\mathrm{eff}}(r)$ as a function of the radial coordinate for different values of the torsion parameter $\omega$. The presence of the helical geometry introduces an attractive $1/r$ interaction term, which becomes more substantial as $\omega$ increases. This leads to the formation of a deeper potential well for larger torsion, thus favoring the existence of bound states. The effective potential results from the combination of the modified centrifugal barrier, proportional to $1/r^2$, and the torsion-induced attractive component proportional to $1/r$. The enhancement of the attractive part is a direct consequence of the coupling between the longitudinal and azimuthal degrees of freedom present in the helically twisted space. 

The physical meaning of each term becomes transparent: the first term proportional to $E$ carries the quantum energy dependence; the second term renormalizes the longitudinal momentum contribution due to torsion; the third term represents the modified centrifugal barrier; while the last term introduces a new effective potential proportional to $1/r$, arising from the helical torsion-induced coupling between angular and longitudinal momenta. This last term has no analog in flat space and reflects the intrinsic geometric twisting of the space.

The structure of this equation reveals several significant consequences for quantum dynamics. The parameter $\omega$ modifies the effective mass along the $z$-direction, increases the strength of the centrifugal barrier, and induces a coupling between $m$ and $k$. In particular, the appearance of a $1/r$ term recalls the form of spin-orbit coupling potentials, though here it emerges purely from geometric torsion.

In the next section, we obtain the exact solution and energy spectrum under the influence of the helically twisted spacetime.

\section{Exact Solution and Energy Spectrum \label{solutions}}

Now, from Eq. \eqref{radial_equation_expanded}, we consider the radial equation for bound states, written in the form
\begin{align}
- \frac{\hbar^2}{2\mu} \Bigg[ \frac{d^2 f(r)}{dr^2} &- \frac{m^2 - 1/4}{r^2} f(r) + \frac{2km\omega}{r} f(r) \notag\\&- k^2 (1+\omega^2) f(r) \Bigg] = \kappa^2 f(r),
\end{align}
where $\kappa = \sqrt{-E}>0$ to ensure that we seek bound energies states. This differential equation admits closed-form solutions in terms of the confluent hypergeometric function ${}_1F_1$. The physical (regular) solution reads
\begin{equation}
f(r) = C_n \, r^{\frac{1}{2} + |m|} e^{-\rho r} \, {}_1F_1\left(a, b, 2\rho r\right),
\end{equation}
where $C_n$ is the normalization constant, and
\begin{align}
\rho &= \frac{1}{\hbar}\sqrt{-2\mu E + \hbar^2 k^2(1+\omega^2)}, \\
a &= \frac{1}{2} + |m| - \frac{k m \omega \hbar}{\rho}, \\
b &= 1 + 2|m|.
\end{align}

The bound state condition requires that the confluent hypergeometric series be truncated into a polynomial, imposing the quantization rule
\begin{equation}
a = -n, \quad \text{with} \quad n \in \mathbb{N}_0.
\end{equation}
Thus, we find the relation
\begin{equation}
\frac{k m \omega \hbar}{\rho} = \frac{1}{2} + |m| + n. \label{rel}
\end{equation}
By solving (\ref{rel}) for $E$, we obtain the exact energy spectrum
\begin{equation}
E_{n,m} = -\frac{k^2 \hbar ^2}{2 \mu}\frac{4 m^2 + (1+2n)(1+\omega^2)(1 + 2n + 4|m|)}{(1+2n+2|m|)^2}.\label{energies}
\end{equation}
\begin{figure}[t!]
\centering
\includegraphics[scale=0.41]{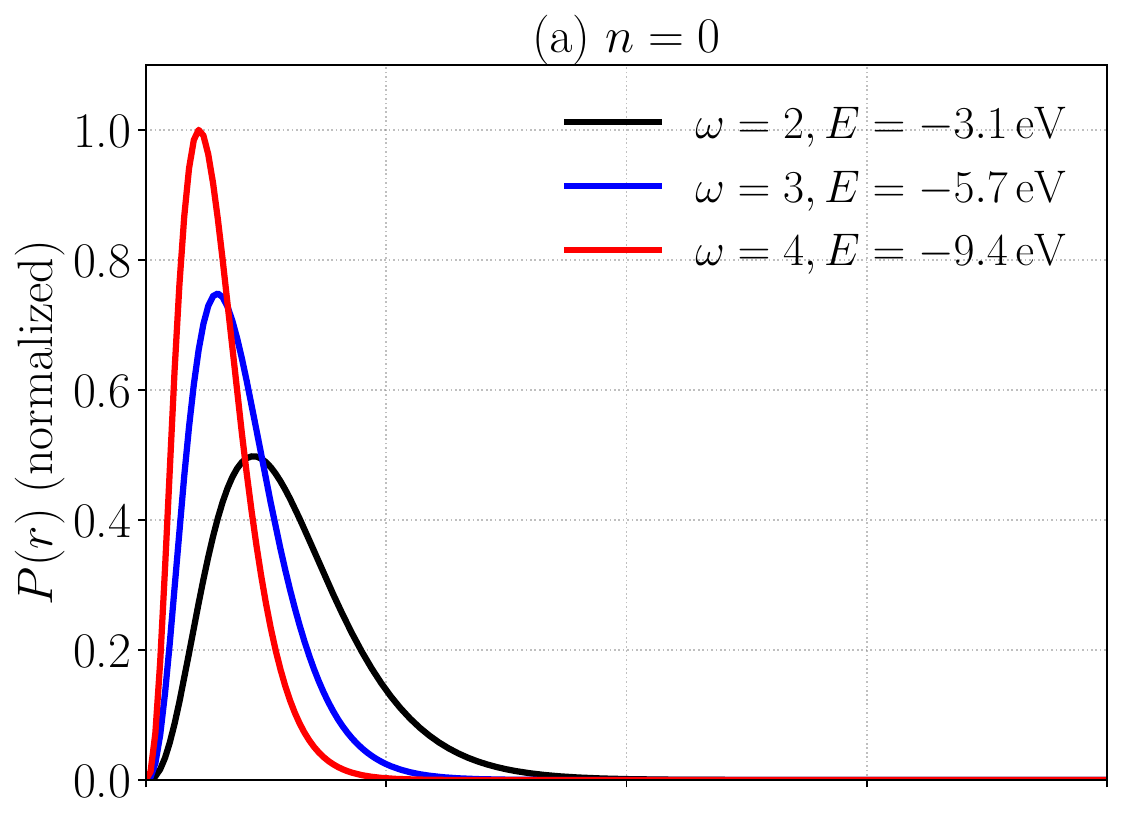}
\includegraphics[scale=0.41]{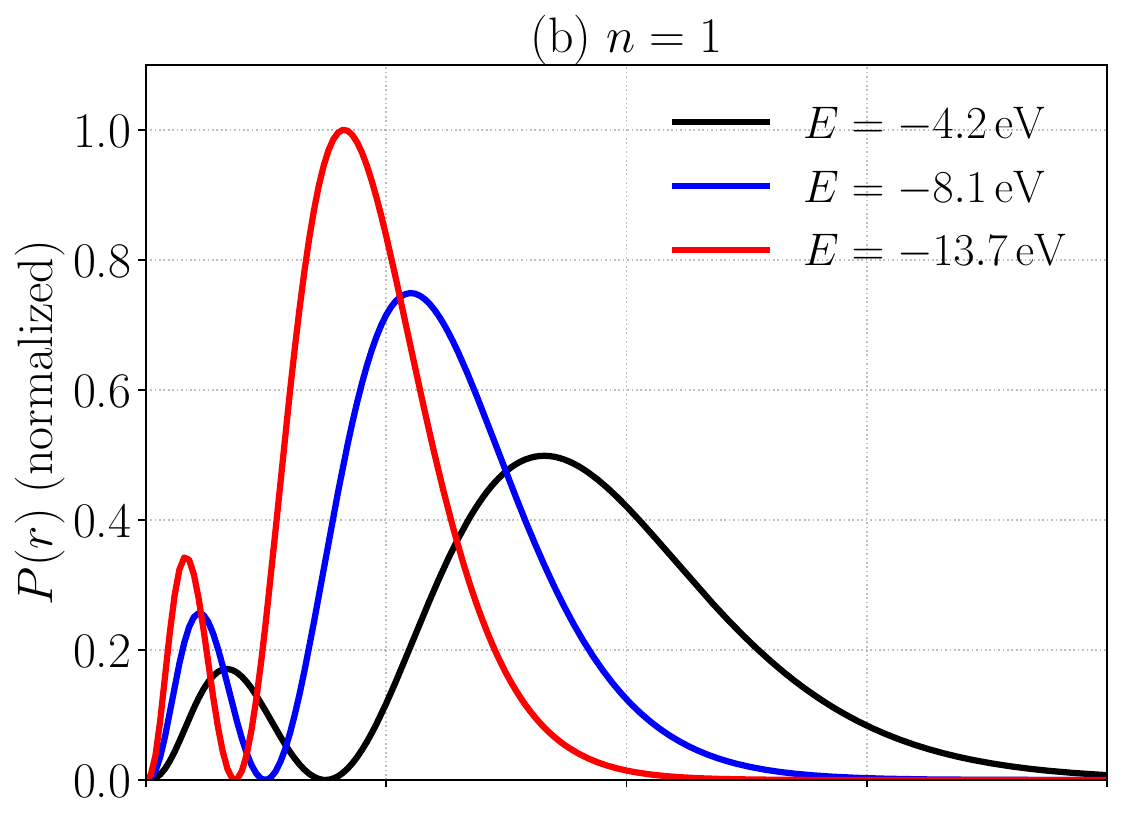}
\includegraphics[scale=0.41]{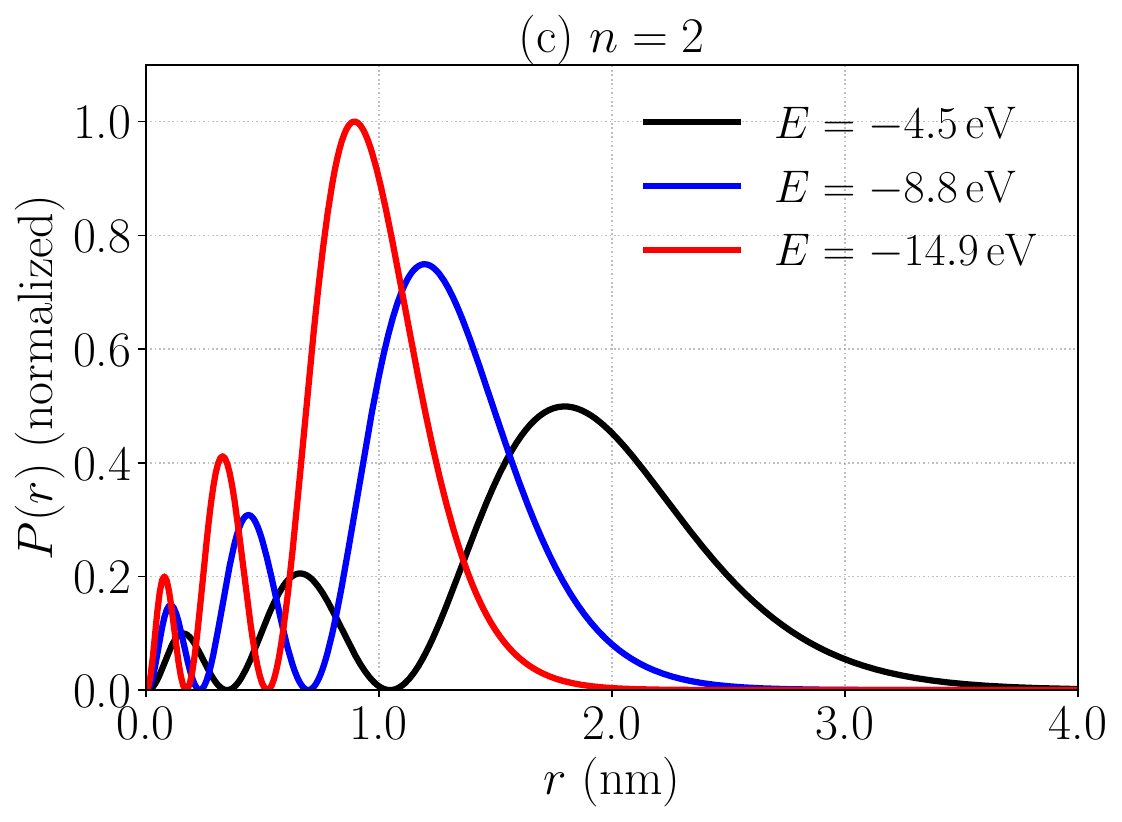}
\caption{\footnotesize Normalized radial probability densities $P(r)$ for the bound states $n=0$,  $n=1$ and $n=2$ as a function of the radial coordinate $r$, for three values of the torsion parameter $\omega = 0.3$, $0.5$, and $0.7$. In Fig. (a), corresponding to the ground state ($n=0$), the probability density becomes increasingly localized near the origin as $\omega$ increases, reflecting the stronger effective attraction. In Fig. (b), corresponding to the first excited state ($n=1$), the probability densities exhibit broader spatial extension with a single node, and similarly shift toward smaller $r$ values for larger $\omega$. In Fig. (c), the second excited state ($n = 2$) is shown, exhibiting a greater number of nodes. We use $m=1$ and $k_{z}=5.0\times10^{9}$ m$^{-1}$.}
\label{fig:Probability}
\end{figure}
All energy levels are strictly negative in this representation. However, since the effective potential tends asymptotically to a positive constant,
\begin{equation}
V_\infty =- \frac{\hbar^2 k^2}{2\mu} (1 + \omega^2),\label{vinf}
\end{equation}
the physical bound states lie in the interval $E_{n,m} \in (-V_\infty, 0)$, as expected for a shifted spectrum. The torsional parameter $\omega$ plays a crucial role in inducing bound states. As $\omega$ increases, it lifts the degeneracy with respect to $m$, introducing a rotational fine-structure in the energy levels. In the limit $\omega \to 0$, the effective Coulomb-like interaction vanishes, and the system reduces to a free particle. Although the analytical expression for $E_{n,m}$ remains formally well-defined when $\omega \to 0$, its physical interpretation must be carefully revisited in this limit. As the torsional parameter approaches zero, the effective potential \eqref{effective_potencial} simplifies to
\begin{equation}
V_{\mathrm{eff}}(r) \Big|_{\omega=0} = -\frac{\hbar^2}{2\mu} \left( -\frac{m^2 - \frac{1}{4}}{r^2} - k^2 \right),
\end{equation}
which corresponds to a purely centrifugal term combined with a constant energy shift. Importantly, no Coulomb-like attractive interaction survives, and the effective potential no longer exhibits a confining well.

Under these conditions, the radial equation reduces to
\begin{equation}
f''(r) + \left[ q^2 - \frac{m^2 - 1/4}{r^2} \right] f(r) = 0,
\end{equation}
where $q= \sqrt{\dfrac{2\mu E}{\hbar^2} - k^2}$. This is the standard Bessel differential equation whose general solution is a linear combination of Bessel functions, given by
\begin{equation}
f(r) = A\, J_{|m|}(qr) + B\, Y_{|m|}(qr).
\end{equation}
Thus, in the absence of torsion, no normalizable bound states exist, and the energy spectrum becomes continuous, corresponding to free-particle solutions with cylindrical symmetry. 

The discrete spectrum obtained analytically for $\omega \neq 0$ emerges only due to the geometric torsion, which induces a Coulomb-like potential capable of binding the particle. Therefore, while the expression for $E_{n,m}$ is mathematically smooth as $\omega \to 0$, it does not represent physical bound states in this limit. The transition $\omega \to 0$ corresponds to a singular limit from the point of view of the bound-state problem, where the very existence of bound states requires nonzero torsion.

Figure~\ref{fig:Probability} illustrates how geometric torsion modifies the spatial distribution of the bound‐state wavefunctions for the first three radial modes. In both figures we use $m=1$ and $k_{z}=5.0\times10^{9}$ m$^{-1}$. In panel (a) ($n=0$), one sees that increasing the torsion parameter $\omega$ shifts the peak of the ground‐state probability density to smaller radii and simultaneously raises its maximum, indicating stronger radial confinement. Panel (b) ($n=1$) shows the same trend for the first excited state: both the principal maximum and the nodal structure move inward as $\omega$ increases. Finally, in panel (c) ($n=2$) the two nodes of the second excited state draw closer together and the overall envelope of $P(r)$ contracts toward the origin with larger $\omega$. This systematic inward migration and sharpening of all probability peaks directly mirror the deepening and narrowing of the Coulomb‐like well induced by the helical torsion in the effective potential.  

These results clearly demonstrate how the combined effects of curvature and torsion deeply influence both the structure of the wavefunctions and the quantization of energy levels in helically twisted spaces.
\begin{figure}[tbhp]
\centering
\includegraphics[scale=0.43]{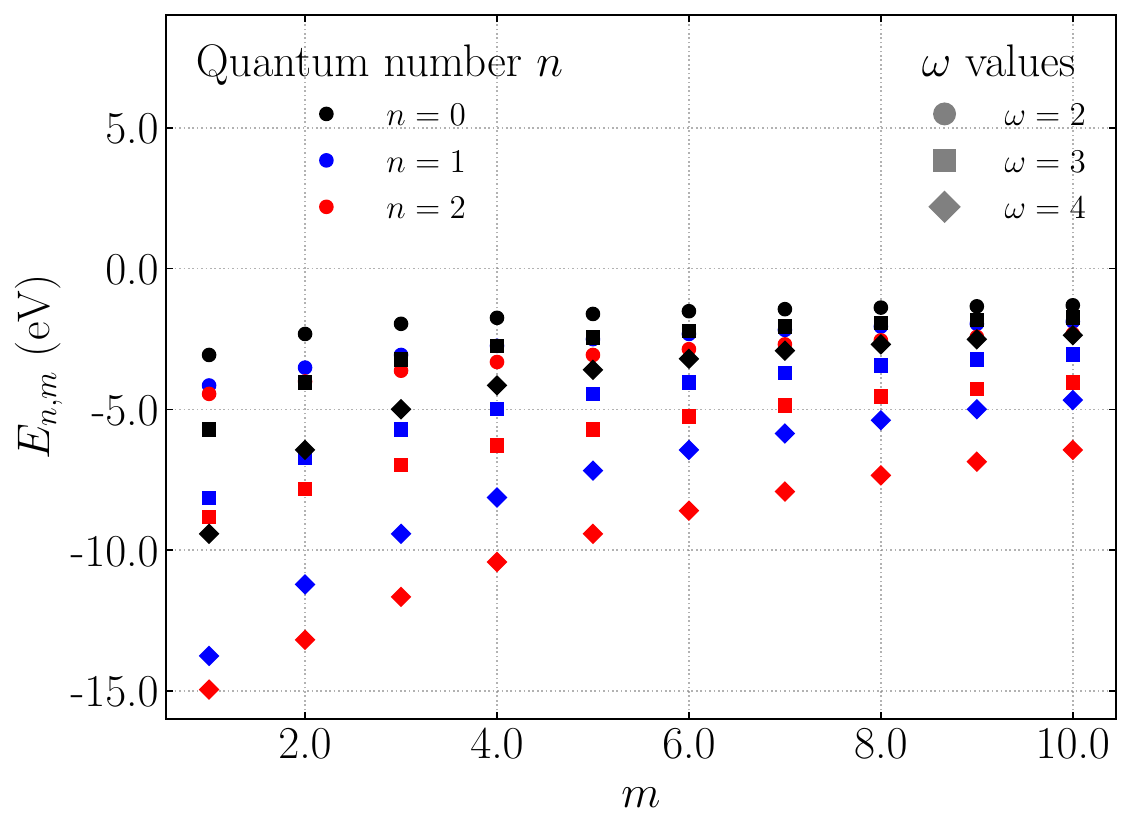}
\caption{\footnotesize Energy spectrum $E_{n,m}$ as a function of the azimuthal quantum number $m$ for $n=0$ (black), $n=1$ (blue), and $n=2$ (red), considering three different values of the torsion parameter $\omega = 2.0$, $3.0$, and $4.0$. The torsion induces a coupling between $m$ and $k$, leading to a shift and splitting of the energy levels as $\omega$ increases. The degeneracy present in the flat case is completely lifted due to the torsional interaction. We use $k_{z}=5.0\times10^{9}$ m$^{-1}$.}
\label{fig:Energy_vs_m}
\end{figure}
\begin{figure}[ht]
\centering
\includegraphics[scale=0.43]{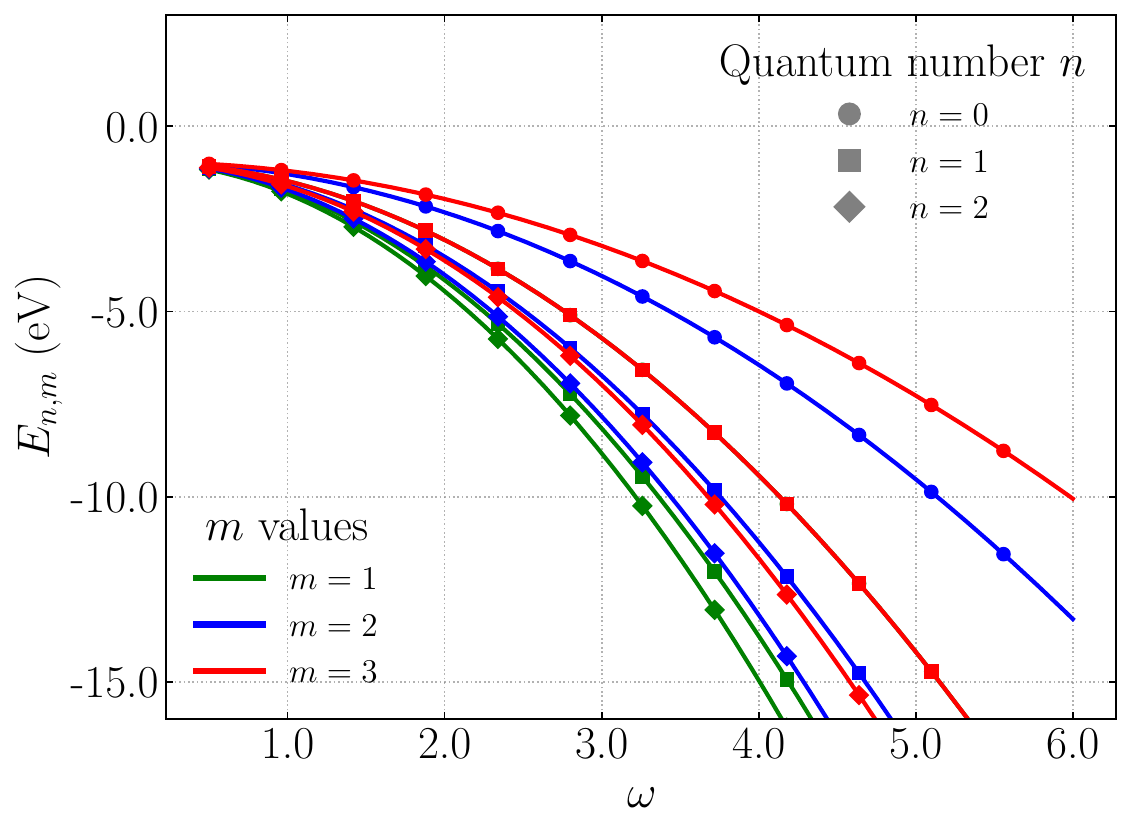}
\caption{Energies $E_{n,m}$ (in~eV) obtained from Eq.~(\ref{energies}) as a function of the rotational frequency $\omega$ (dimensionless) for the helically twisted spacetime model.  The three radial quantum numbers $n = 0, 1, 2$ are identified by distinct markers (circles, squares, and diamonds), while the three orbital indices $m = 1, 2, 3$ are colour–coded as green, blue, and red, respectively. We use $k_{z}=5.0\times10^{9}$ m$^{-1}$.}
\label{fig:Enm_vs_omega}
\end{figure}

Figure~\ref{fig:Energy_vs_m} shows the dependence of the energy levels $E_{n,m}$ on the azimuthal quantum number $m$ for the first three radial modes ($n=0, 1, 2$) and for three representative values of the torsion parameter $\omega$. As expected, the torsional interaction significantly modifies the energy spectrum, producing a nonlinear behavior of $E_{n,m}$ as a function of $m$. The coupling between the angular and longitudinal momenta, mediated by $\omega$, removes the degeneracy that would otherwise exist in the absence of torsion. The energy shifts become more pronounced for larger values of $\omega$, especially for higher values of $m$, where the effective torsional coupling is stronger.

The spectrum displayed in Fig. \ref{fig:Enm_vs_omega} exhibits three salient trends.  
(i) For every fixed pair $(n,m)$, the energy drops monotonically with the rotational frequency~$\omega$, showing that the centrifugal contribution introduced by rotation systematically lowers the bound levels.  
(ii) For a given radial index $n$ the states with larger azimuthal index $m$-drawn in green ($m=1$), blue ($m=2$), and red ($m=3$) remain energetically lower than those with smaller $m$ across the whole $\omega$ interval, indicating that the $\omega$--dependent centrifugal shift outweighs the static $m$--dependent barrier already present at $\omega = 0$.  
(iii) The separation between successive radial manifolds (markers $\circ$, $\square$, $\diamond$ for $n = 0, 1, 2$) shrinks as $\omega$ increases, hinting at an incipient clustering of levels that could enable resonant inter-manifold transitions in the high-rotation regime.

Quantitatively, the crossover from weak to moderate rotation ($\omega \simeq 0.5$-$3$) produces shifts of a few~eV--well within the resolution of optical and magneto-transport probes in semiconductor quantum rings.  
In the strong-rotation domain ($\omega \gtrsim 5$) the near-degeneracy of different $n$ manifolds, together with the persistent splitting among $m$ branches, is expected to enhance the amplitude of persistent currents through inter-level mixing.  
Altogether, Fig.~\ref{fig:Enm_vs_omega} offers a compact visual account of how the radial ($n$) and azimuthal ($m$) quantum numbers, in concert with rotation, sculpt the bound-state spectrum of the helically twisted spacetime model.

\begin{table}[tbhp]
\centering
\caption{\footnotesize Comparison between the numerical ($E_{\mathrm{num.}}$) and analytical ($E_{\mathrm{analyt.}}$) bound-state energies for different values of the torsion parameter $\omega$, radial quantum number $n$, and azimuthal quantum number $m$. The excellent agreement demonstrates the full consistency between the finite difference method and the exact analytical solution.}
\label{tab:energy_comparison}
\setlength{\tabcolsep}{4pt}
\renewcommand{\arraystretch}{1.1}
\vspace{0.2cm}
\begin{tabular}{cccccc}
\toprule
$\omega$ & $n$ & $m$ & $E_{\mathrm{num.}}$(eV) & $E_{\mathrm{analyt.}}$(eV) & $\Delta E$ (eV) \\
\midrule
2 & 0 & 1 & -3.0652 & -3.0692 & $4.0000 \times 10^{-3}$ \\
2 & 1 & 1 & -4.1519 & -4.1529 & $1.0000 \times 10^{-3}$ \\
2 & 2 & 1 & -4.4511 & -4.4515 & $4.0000 \times 10^{-4}$ \\
2 & 0 & 2 & -2.3236 & -2.3241 & $5.0000 \times 10^{-4}$ \\
2 & 1 & 2 & -3.5181 & -3.5184 & $3.0000 \times 10^{-4}$ \\
2 & 2 & 2 & -4.0097 & -4.0099 & $2.0000 \times 10^{-4}$ \\
2 & 0 & 3 & -1.9631 & -1.9633 & $2.0000 \times 10^{-4}$ \\
2 & 1 & 3 & -3.0689 & -3.0692 & $3.0000 \times 10^{-4}$ \\
2 & 2 & 3 & -3.6287 & -3.6289 & $2.0000 \times 10^{-4}$ \\
3 & 0 & 1 & -5.6969 & -5.7150 & $1.8100 \times 10^{-2}$ \\
3 & 1 & 1 & -8.1487 & -8.1534 & $4.7000 \times 10^{-3}$ \\
3 & 2 & 1 & -8.8233 & -8.8252 & $1.9000 \times 10^{-3}$ \\
3 & 0 & 2 & -4.0361 & -4.0386 & $2.5000 \times 10^{-3}$ \\
3 & 1 & 2 & -6.7240 & -6.7258 & $1.8000 \times 10^{-3}$ \\
3 & 2 & 2 & -7.8305 & -7.8316 & $1.1000 \times 10^{-3}$ \\
3 & 0 & 3 & -3.2257 & -3.2268 & $1.1000 \times 10^{-3}$ \\
3 & 1 & 3 & -5.7137 & -5.7150 & $1.3000 \times 10^{-3}$ \\
3 & 2 & 3 & -6.9734 & -6.9745 & $1.1000 \times 10^{-3}$ \\
4 & 0 & 1 & -9.3662 & -9.4191 & $5.2900 \times 10^{-2}$ \\
4 & 1 & 1 & -13.7403 & -13.7540 & $1.3700 \times 10^{-2}$ \\
4 & 2 & 1 & -14.9430 & -14.9483 & $5.3000 \times 10^{-3}$ \\
4 & 0 & 2 & -6.4311 & -6.4389 & $7.8000 \times 10^{-3}$ \\
4 & 1 & 2 & -11.2106 & -11.2161 & $5.5000 \times 10^{-3}$ \\
4 & 2 & 2 & -13.1785 & -13.1821 & $3.6000 \times 10^{-3}$ \\
4 & 0 & 3 & -4.9923 & -4.9957 & $3.4000 \times 10^{-3}$ \\
4 & 1 & 3 & -9.4151 & -9.4191 & $4.0000 \times 10^{-3}$ \\
4 & 2 & 3 & -11.6547 & -11.6582 & $3.5000 \times 10^{-3}$ \\
\bottomrule
\end{tabular}
\end{table}
\section{Numerical Validation and Accuracy Assessment}
\label{numerical}

To assess the accuracy of the numerical method employed to solve the radial eigenvalue problem, we performed a systematic comparison between the numerical eigenvalues obtained via the finite difference approach and the exact analytical solutions derived in Sec.~\ref{solutions}. The comparison is presented in Table~\ref{tab:energy_comparison}, where we display the bound-state energies for several values of the torsion parameter $\omega$, radial quantum number $n$, and azimuthal quantum number $m$.

As shown in the table, the agreement between the two methods is excellent across all cases examined. The relative deviations $\Delta E_{n,m}$ remain remarkably small, typically of the order of $10^{-3}$~eV or smaller. Even for higher torsion strengths (e.g., $\omega = 4.0$), where the coupling effects become more pronounced, the numerical and analytical results exhibit deviations that do not exceed a few parts in $10^{-2}$~eV. These tiny differences are primarily attributed to the discretization effects inherent in the finite.

\subsection{Symmetry under the sign of the torsional parameter}

It is worth noting that the effective potential $V_{\mathrm{eff}}(r)$ in Eq. (\ref{effective_potencial}) exhibits an interesting symmetry with respect to the sign of the torsional parameter $\omega$. The key term responsible for the existence of bound states is the Coulomb-like contribution proportional to $(2 k m \omega)/r$. When $\omega>0$, attractive wells are formed for positive values of $m$. However, when $\omega<0$, the Coulomb term becomes repulsive for $m>0$, preventing the formation of bound states, and $k$ could be negative. 

Nevertheless, the situation is fully restored for negative values of $m$. Indeed, for $\omega<0$ and $m<0$, the combination $k m \omega$ remains positive, reinstating the attractive character of the potential. Therefore, the bound-state spectrum depends only on the product $m \omega$, and any calculation performed for $\omega>0$ can be directly extended to $\omega<0$ by considering the corresponding states with opposite values of $m$. This symmetry explains why the same energy levels appear for combinations of $(m, \omega)$ and $(-m, -\omega)$.

As a consequence, in the present analysis, it is sufficient to restrict the study to $\omega>0$, since the full spectrum for $\omega<0$ is trivially obtained by exploiting this discrete symmetry. This interplay, involving the role of the sign in $km\omega$, is also observed in the study of the dynamics of massive and massless particles in the spacetime of a wiggly cosmic dislocation~\cite{dos2024dynamics}.

\section{Physical interpretation and further discussions}

Before examining specific cases, we first outline the general physical implications of the torsion-induced effective potential derived in Sec.~V. We highlight its main qualitative features and discuss how geometric twisting alone can generate quantum confinement and modify the bound-state spectrum. We then proceed to compare this purely geometry-driven interaction with the conventional Coulomb potential, analyze the semiclassical limit for highly excited states, and finally sketch potential experimental platforms in which helically twisted quantum systems might be realized.

\subsection{Comparison with conventional Coulomb potential}

Although the effective potential emerging in our model contains a Coulomb-like term proportional to $1/r$, its physical origin differs fundamentally from that of the standard Coulomb potential in atomic systems. In conventional atomic physics, the Coulomb interaction arises from the electromagnetic force between charged particles, described by $V_{\mathrm{Coulomb}}(r) = - Ze^2/r$. Here, instead, the $1/r$ dependence emerges purely from the intrinsic coupling between angular and longitudinal momenta induced by the underlying torsional geometry. In particular, the term proportional to $km\omega/r$ in Eq.~(\ref{effective_potencial}) reflects a purely geometric interaction, arising even in the absence of any external field.

Despite the formal similarity in the mathematical structure of the radial equation, the underlying physics is distinct. The torsion parameter $\omega$ plays the role of an effective coupling constant, analogous to the nuclear charge $Z$ in the Coulomb case. However, in the present context, this coupling can in principle be continuously tuned by adjusting the geometry of the twisted space. The fact that the Coulomb-like behavior arises naturally from the nontrivial spatial structure highlights the richness of geometry-induced quantum interactions.

\subsection{Semiclassical behavior for highly excited states}

It is instructive to analyze the asymptotic behavior of the energy spectrum in the limit of highly excited states, i.e., $n \gg 1$. Starting from the exact expression for the bound-state energies (\ref{energies}), we expand it systematically for large $n$. 
After performing the series expansion to second order in $1/n$, we obtain
\begin{equation}
E_{n\,m} \approx -\frac{k^2 \hbar^2}{2\mu}(1+\omega^2) + \frac{k^2 \hbar^2 \omega^2 m^2}{2\mu} \cdot \frac{1}{n^2} + \mathcal{O}\left( \frac{1}{n^3} \right).
\end{equation}
This result reveals that, for large $n$, the energy spectrum accumulates asymptotically near the constant threshold
\begin{equation}
E_\infty = -\frac{k^2 \hbar^2}{2\mu}(1+\omega^2),
\end{equation}
which corresponds precisely to the shifted continuum threshold associated with the effective potential asymptotic behavior, as stated in Eq. (\ref{vinf}). The second term, proportional to $1/n^2$, describes the residual fine structure in the bound-state spectrum and depends quadratically on the azimuthal quantum number $m$. This structure reflects the residual influence of the torsional parameter $\omega$ on the high-lying bound states. As $n \to \infty$, the bound levels accumulate densely just below the continuum, consistent with general expectations for confining potentials of Coulomb-like nature.

It is worth emphasizing that, unlike conventional Coulomb systems where the energy levels vanish as $E \sim 1/n^2$, here the torsion-induced shift produces a nonzero asymptotic offset $E_\infty$, intrinsic to the underlying helical geometry. 

\subsection{Possible physical realizations}

Although our study focuses on the theoretical aspects of quantum dynamics in helically twisted spaces, it is worthwhile to discuss potential physical scenarios where such models might find relevance. Recent advances in the fabrication of artificial helical nanostructures, twisted nanoribbons, and curved quantum waveguides suggest possible condensed matter realizations of the present model \cite{IainJClark_1996,PhysRevB.91.245412}. In such systems, effective torsional couplings may arise from lattice distortions, dislocation defects, or engineered curvature at the nanoscale. Additionally, twisted bilayer graphene structures, where moiré superlattices are formed by rotational stacking, provide platforms where interlayer coupling leads to rich topological and geometric effects \cite{cao2018correlated}. While our helically twisted metric offers a simplified toy model, the underlying mechanism of geometry-induced confinement may find concrete realization in engineered materials where torsion-like couplings can be introduced via external strain, defects, or controlled twisting.

\section{Harmonic oscillator system}

Before applying this framework to the harmonic oscillator, it is instructive to highlight the significance of this model. The quantum harmonic oscillator remains one of the cornerstones of theoretical and applied physics, serving not only as an exactly solvable system that underpins vibrational modes in molecules and solid-state phonons, but also as the foundation for perturbation theory and quantum field quantization. Extending the oscillator to curved or twisted geometries reveals how spatial deformations modify confinement and energy quantization, with potential implications for engineered nano-mechanical resonators, optical waveguides in helical fibers, and quantum devices subject to strain or torsion.

The harmonic oscillator system in curved space using the Laplace-Beltrami operator can be written as \cite{LDL}
\begin{equation}
E\,\Psi ({\bf r})=\left[-\frac{\hbar^2}{2\mu}\frac{1}{\sqrt{g}} \partial_i \left( \sqrt{g}g^{ij}\partial_j\right)+\frac{1}{2}\,\mu\,\omega^2_0\,r^2\right]\,\Psi ({\bf r}), \label{aa1}
\end{equation}
where $\omega_0$ is the oscillator frequency, $\vec{r}$ is the distance of the particle from the origin, and others are mentioned earlier. 

Using the given metric components, we can re-write Eq. (\ref{aa1}) explicitly as,
\begin{align}
E\Psi ({\bf r}) &= \Bigg\{-\frac{\hbar^2}{2\mu} \Bigg[ 
\frac{1}{r} \partial_r \left( r \partial_r\right) 
+ \frac{\partial_\phi^2}{r^2} 
- \frac{2\,\omega}{r} \partial_\phi\, \partial_z \notag\\ &+ (1+\omega^2) \partial_z^2\Bigg]+\frac{1}{2}\,\mu\,\omega^2_0\,r^2\Bigg\}\,\Psi ({\bf r}). \label{aa2}
\end{align}

Using the wave-function ansatz given in Eq. (\ref{ansatz}), we find the following equation:
\begin{align}
E\,\psi (r) &= \Bigg\{-\frac{\hbar^2}{2\mu} \Bigg[ 
\frac{1}{r} \partial_r \left( r\, \partial_r\right) 
- \frac{m^2}{r^2} 
+\frac{2\,m\,\omega\,k}{r}\notag\\ &-(1+\omega^2)\,k^2\Bigg]+\frac{1}{2}\,\mu\,\omega^2_0\,r^2\Bigg\}\,\psi (r). \label{aa3}
\end{align}

After applying the transformation $ \psi(r) = r^{-1/2}\, \mathrm{R}(r) $, which eliminates the first derivative, the radial equation takes the form
\begin{equation}
-\frac{\hbar^2}{2\mu}\,\mathrm{R}''(r) + V_{\mathrm{eff}}(r)\,\mathrm{R}(r) = E\,\mathrm{R}(r),\label{aa4}
\end{equation}
where the effective potential $ V_{\mathrm{eff}}(r) $ is given by
\begin{align}
V_{\mathrm{eff}}(r) = -\frac{\hbar^2}{2\mu} \Bigg[ - \frac{m^2-\frac{1}{4} }{r^2} &+ \frac{2\,\omega\, k\, m}{r} - (1+\omega^2)\,k^2\notag\\&-\Omega^2\,r^2 \Bigg],\label{aa5}
\end{align}
where $\Omega=|\mu\,\omega_0/\hbar|$. 

To solve Eq. (\ref{aa4}), we can rewrite this equation as follows:
\begin{equation}
   \mathrm{R}''(r)+\left[\Lambda^2-\Omega^2\,r^2-\frac{\iota^2}{r^2}+\frac{2\,\xi}{r}\right]\,\mathrm{R}(r)=0,\label{aa6} 
\end{equation}
where we set the following parameters
\begin{align}
    \Lambda &=\sqrt{\frac{2\,\mu\,E}{\hbar^2}-(1+\omega^2)\,k^2},\nonumber\\
    \iota &=\sqrt{m^2-1/4}>0,\nonumber\\
    \xi &=m\,\omega\,k.\label{aa7}
\end{align}
Special functions such as the (confluent) Heun function can, in principle, describe the general solution of the radial equation.  In practice, however, Heun-class equations are more robustly treated by finite difference discretization. High-order schemes, chiefly the classic and matrix versions of the Numerov algorithm, replace the differential operator by a sparse band matrix whose diagonalization yields the full bound-state spectrum and the associated
eigenfunctions, free of the convergence and truncation \cite{AJP.2012.80.1017,Angraini2018NSFDTD,Springel2011Numerov}.

\begin{figure}[htbp]
\centering
\includegraphics[scale=0.43]{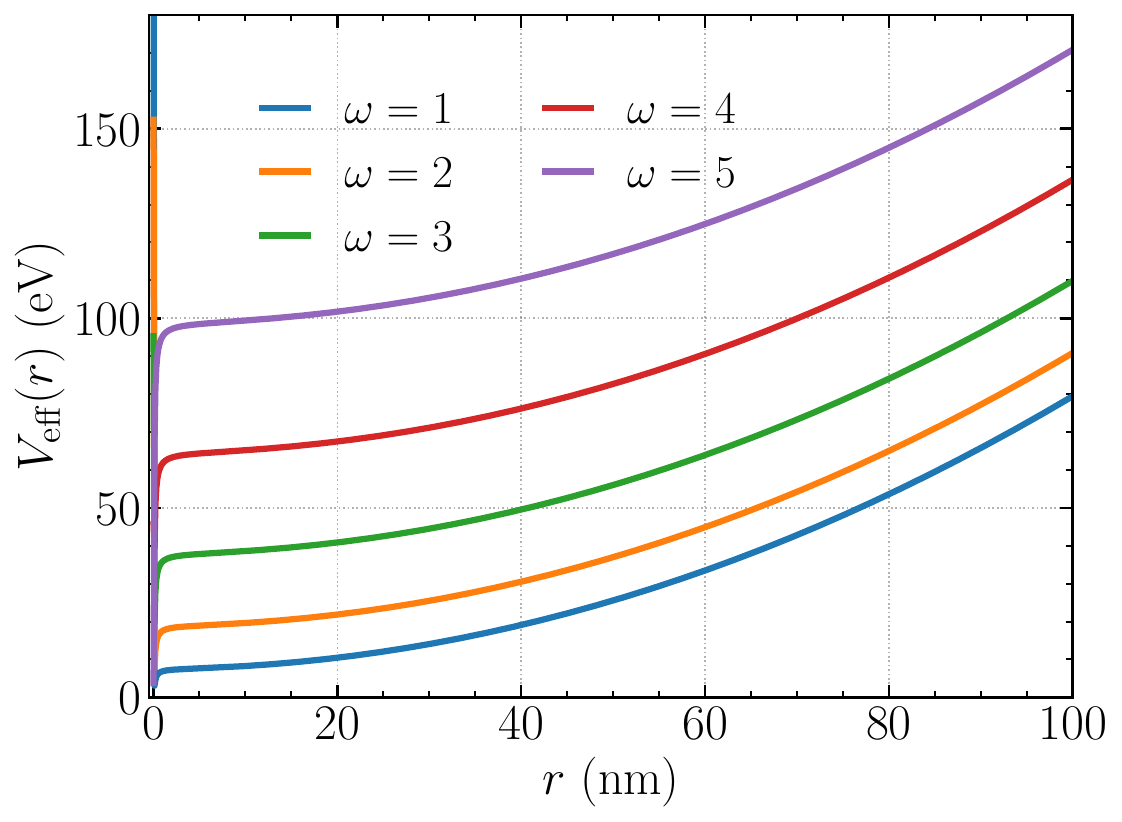}
\caption{\footnotesize Effective potential $V_{\mathrm{eff}}(r)$ (Eq. (\ref{aa5})) plotted as a function of the radial coordinate $r$ (in nm) for several values of the dimensionless torsion parameter $\omega\in\{1,2,3,4,5\}$. The curves exhibit a pronounced centrifugal barrier at small $r$, whose height and position shift upward and outward as $\omega$ increases.}
\label{fig:Veff_twist}
\end{figure}

Let us consider the following radial function
\begin{equation}
    \mathrm{R}(r)=r^{\iota}\,e^{-\frac{1}{2}\,\Omega\,r^2}\,e^{-\beta\,r}\,H(r),\label{aa8}
\end{equation}
where $H(r)$ and $\beta$ are unknown to be determined.
\begin{figure}[tbhp]
\centering
\includegraphics[scale=0.43]{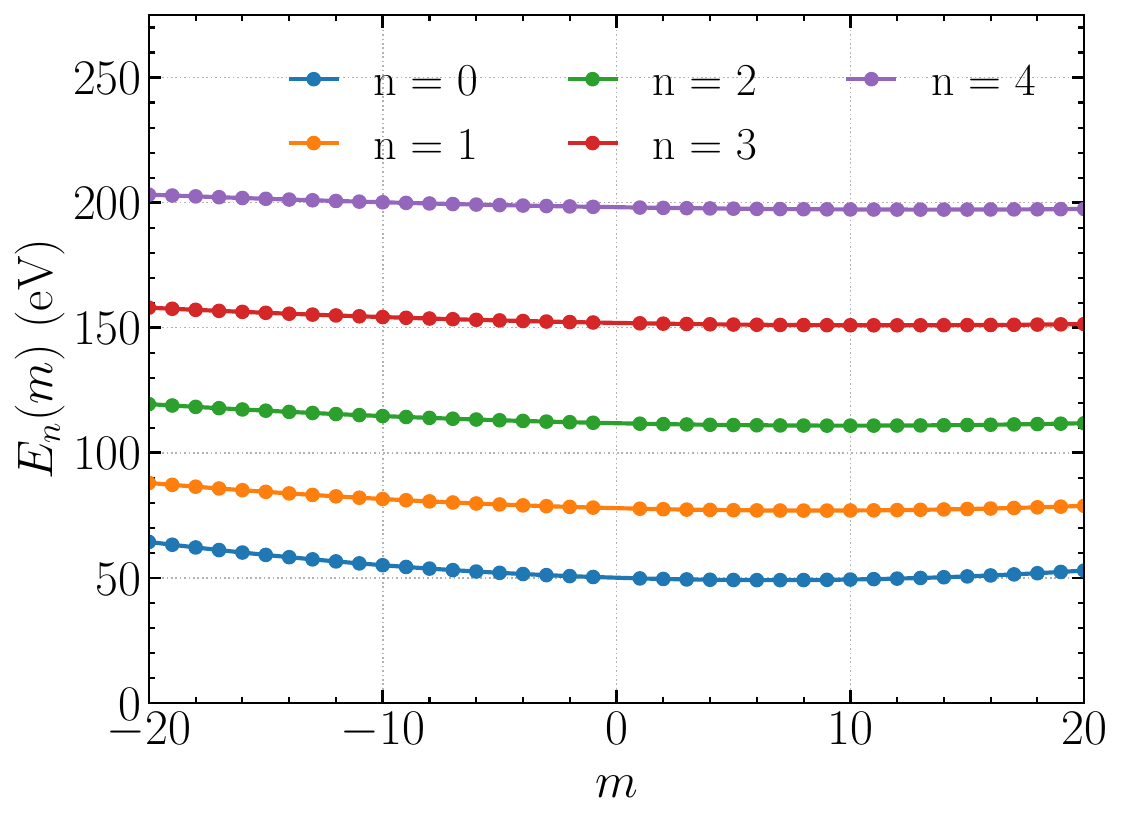}
\caption{\footnotesize Low–lying energy levels $E_{n}(m)$ of a twisted two-dimensional quantum oscillator as a function of the angular‐momentum quantum number $m$.  Parameters are $\omega_{0}=2\pi\times5\times10^{14}\,\text{rad\,s}^{-1}$, torsion strength $\omega=5$, longitudinal wave number $k_{z}=10^{9}\,\text{m}^{-1}$ and effective mass $\mu=m_{e}$.  The first five radial modes ($n=0,1,2,3,4$) were obtained by finite-difference diagonalization with Dirichlet boundaries; successive points are connected by a state-following algorithm that maximizes the overlap between neighboring eigenvectors, ensuring a smooth evolution through quasi-crossings.}
\label{fig:Enm_twist5}
\end{figure}
\begin{figure}[tbhp]
\centering
\includegraphics[scale=0.43]{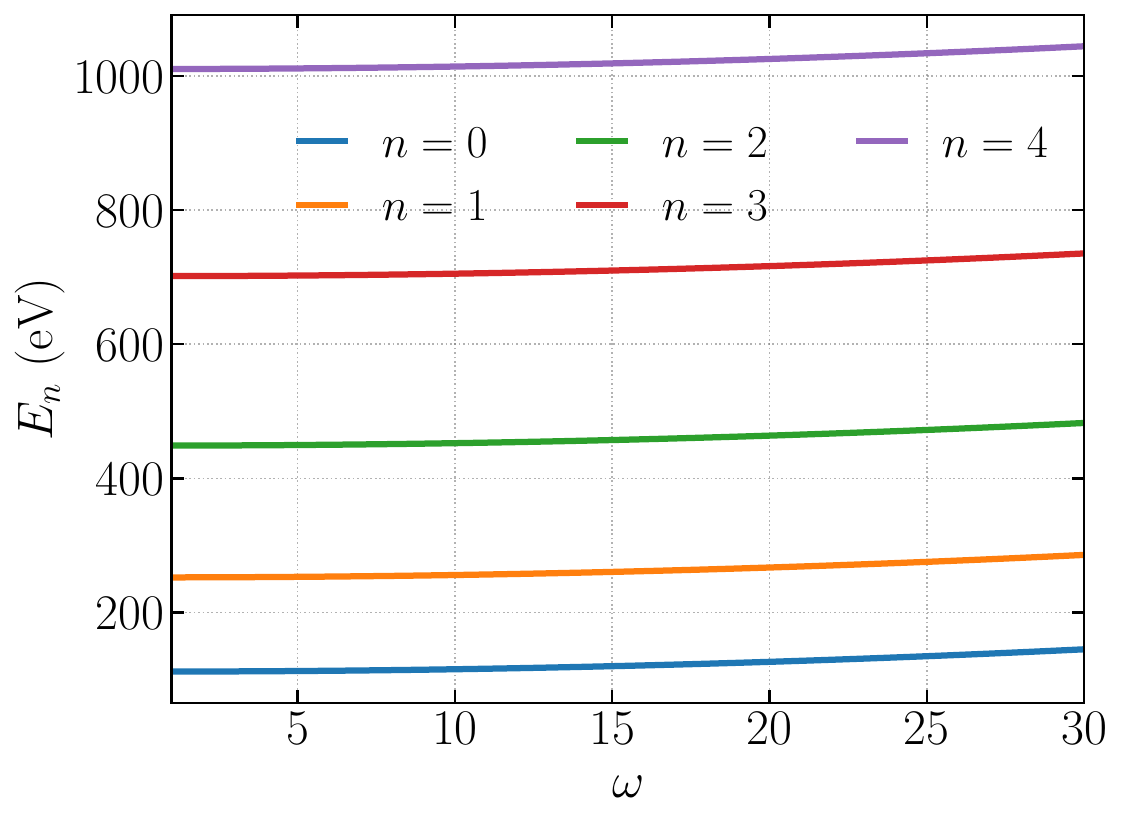}
\caption{\footnotesize Lowest five radial energies $E_{n}$ as functions of the torsion parameter $\omega$ for a fixed angular momentum $m=1$. The oscillator frequency is $\omega_{0}=2\pi\times5\times10^{14}\,\text{rad\,s}^{-1}$
and the longitudinal wave-number is $k_{z}=10^{9}\,\text{m}^{-1}$. All curves rise monotonically because of the positive constant shift $E_{\text{shift}}=\hbar^{2}(1+\omega^{2})k_{z}^{2}/(2\mu)$ grows quadratically with $\omega$ and dominates over the linear coupling term that scales as $-\,\omega/r$. The upward trend is stronger for higher radial quantum numbers, reflecting their larger expectation value of $r^{2}$ and therefore a greater sensitivity to the torsion-induced shift.}
\label{fig:En_twist}
\end{figure}
\begin{table*}[tbhp]
  \centering
  \caption{\footnotesize Energy levels $E_n$ (eV) as functions of the torsion parameter $\omega$, for the first five radial modes at 25 representative $\omega$ values. These data correspond to the curves displayed in Fig.~\ref{fig:En_twist}.}
  \label{tab:E_vs_twist_25}
  \setlength{\tabcolsep}{4pt}
  \renewcommand{\arraystretch}{1.1}
  \vspace{0.2cm}
  \begin{tabular}{r | r r r r r}
    \toprule
    $\omega$ & $E_{n=0}$ & $E_{n=1}$ & $E_{n=2}$ & $E_{n=3}$ & $E_{n=4}$ \\
    \midrule
    1.000000 & 112.1989 & 252.4949 & 448.9021 & 701.4224 & 1010.0567 \\
    1.591837 & 112.2348 & 252.5383 & 448.9493 & 701.4718 & 1010.1076 \\
    2.183673 & 112.2974 & 252.6084 & 449.0231 & 701.5480 & 1010.1853 \\
    2.775510 & 112.3866 & 252.7052 & 449.1237 & 701.6507 & 1010.2896 \\
    3.367347 & 112.5026 & 252.8287 & 449.2509 & 701.7802 & 1010.4205 \\
    3.959184 & 112.6451 & 252.9780 & 449.4054 & 701.9368 & 1010.5771 \\
    4.551020 & 112.8143 & 253.1533 & 449.5871 & 702.1204 & 1010.7601 \\
    5.142857 & 113.0101 & 253.3534 & 449.7960 & 702.3310 & 1010.9705 \\
    5.734694 & 113.2326 & 253.5779 & 450.0320 & 702.5685 & 1011.2082 \\
    6.326531 & 113.4816 & 253.8264 & 450.2947 & 702.8327 & 1011.4731 \\
    6.918367 & 113.7569 & 254.0984 & 450.5836 & 703.1237 & 1011.7653 \\
    7.510204 & 114.0585 & 254.3936 & 450.8983 & 703.4413 & 1012.0847 \\
    8.102041 & 114.3863 & 254.7117 & 451.2383 & 703.7854 & 1012.4314 \\
    8.693878 & 114.7400 & 255.0522 & 451.6033 & 704.1546 & 1012.8053 \\
    9.285714 & 115.1200 & 255.4148 & 451.9928 & 704.5487 & 1013.2064 \\
    9.877551 & 115.5260 & 255.7989 & 452.4063 & 704.9673 & 1013.6347 \\
   10.469388 & 115.9580 & 256.2042 & 452.8435 & 705.4099 & 1014.0901 \\
   11.061224 & 116.4157 & 256.6312 & 453.3040 & 705.8763 & 1014.5727 \\
   11.653061 & 116.8991 & 257.0798 & 453.7873 & 706.3662 & 1015.0824 \\
   12.244898 & 117.4080 & 257.5506 & 454.2939 & 706.8803 & 1015.6193 \\
   12.836735 & 117.9423 & 258.0445 & 454.8234 & 707.4183 & 1016.1834 \\
   13.428571 & 118.5018 & 258.5621 & 455.3755 & 707.9799 & 1016.7745 \\
   14.020408 & 119.0863 & 259.1043 & 455.9508 & 708.5648 & 1017.3928 \\
   14.612245 & 119.6958 & 259.6710 & 456.5489 & 709.1727 & 1018.0382 \\
   15.204082 & 120.3301 & 260.2631 & 457.1705 & 709.8032 & 1018.7107 \\
    \bottomrule
  \end{tabular}
\end{table*}
In Figure~\ref{fig:Veff_twist} we display the full effective potential (\ref{aa5})
for a fixed angular quantum number $m=1$, the oscillator frequency
$\omega_0=2\pi\times8\times10^{12}\,\mathrm{rad/s}$, and longitudinal
momentum $k_z=10^{10}\,\mathrm{m}^{-1}$.  Each curve corresponds to a
different torsion $\omega$, illustrating how the linear $1/r$ term
and constant shift $\propto1+\omega^2$ progressively raise the
barrier height and pushing its effective radius outward, dominating the
purely harmonic $\Omega^2 r^2$ confinement at small $r$.  This
behavior underlies the modification of the bound‐state spectrum in the
helicoidal metric.

Figure~\ref{fig:Enm_twist5} illustrates how the bound-state spectrum reacts to a moderate torsional deformation.  For every radial quantum number the curve exhibits a broad minimum around $m\simeq0$ and rises almost symmetrically towards large positive or negative $|m|$, reflecting the competition between the harmonic confinement, the centrifugal barrier and the linear torsion–induced coupling $\propto m \omega k_{z}/r$.  Because $\omega=5$ is still small compared with the natural oscillator scale, the torsional term shifts the entire spectrum upward without introducing a pronounced $m\leftrightarrow -m$ asymmetry; nevertheless, the slight left–right imbalance visible at high $|m|$ signals the breaking of the exact degeneracy that would hold in the untwisted case.  As $n$ increases, the centrifugal contribution weakens, so higher modes sit closer together and their parabolic envelopes flatten. This trend will prove relevant when discussing optical selection rules driven by transitions between different $m$ branches.

Figure~\ref{fig:En_twist} shows that increasing the torsional deformation steadily elevates the entire spectrum. Between $\omega=1$ and $\omega=30$ the ground state rises from $112.2$\,eV to $145.3$\,eV, while the fourth-excited level ($n=4$) climbs from $1010.1$\,eV to $1043.9$\,eV. The nearly linear appearance over this interval reflects the fact that the quadratic shift $\propto(1+\omega^2)k_z^2/(2\mu)$ is being sampled on a restricted range; extending the scan to larger $\omega$ would recover the expected parabolic curvature.  
 
Table~\ref{tab:E_vs_twist_25} presents a sample of the computed bound‐state energies that correspond to the spectral curves plotted in Fig.~\ref{fig:En_twist}. For twenty-five equally spaced values of the dimensionless torsion parameter $\omega\in[1,30]$, the first five radial eigenvalues $E_n$ (in eV) are listed, as obtained by finite-difference diagonalisation of the effective Hamiltonian. As $\omega$ increases, each energy level undergoes a clear upward shift, a direct consequence of the growing constant term $(1+\omega^2)k_z^2/(2\mu)$ in the effective potential. The ground state $E_0$ exhibits the gentlest slope, whereas higher excited modes $E_{n>0}$ climb more steeply, reflecting their larger radial extent and enhanced sensitivity to the torsion-induced $1/r$ coupling. This concise tabulation not only confirms the monotonic trends visible in Fig.~\ref{fig:En_twist}, but also provides precise numerical benchmarks for future analytical or experimental comparisons.

\section{Conclusions \label{conclusion}}

In this article, we have investigated the Schrödinger wave equation in a three-dimensional helically twisted spacetime. We derived the radial equation governing quantum dynamics in this unique geometry, highlighting the geometric coupling between the angular and translational quantum numbers that introduces an attractive Coulomb-like potential solely due to spatial structure. Exact analytical solutions for the bound states, including the energy spectra and normalized wave functions, are obtained. We have demonstrated that the presence of helical torsion, represented by the parameter $\omega$, significantly alters the effective potential in the corresponding wave equation of the quantum system. This geometric deformation introduces non-trivial modifications to the system's dynamics, showing that geometry alone-even in the absence of external fields or interactions-can lead to quantum confinement mechanisms and support the emergence of bound states. Specifically, we analyzed how the helical torsion affects the spectral properties of the system, showing that the eigenvalue solutions differ from those obtained in the standard Minkowski flat space. This highlights the role of torsional geometry in shaping quantum behaviour and enabling geometrically induced confinement mechanisms.

Our findings demonstrate the richness of quantum physics in curved spaces, illustrating how geometric deformations can induce effective interactions and confinement even in the absence of external fields. Helical torsion emerges as pivotal in generating attractive potentials and quantized energy levels, with potential applications in condensed matter systems, quantum materials, and cosmological models. 

Before concluding, we note that this model can be extended to relativistic particles, which would further advance our understanding of geometric effects in quantum mechanics. This extension is currently under investigation and will be reported in future work.

\section*{Acknowledgments}

This work was supported by CAPES (Finance Code 001), CNPq (Grant 306308/2022-3), and FAPEMA (Grants UNIVERSAL-06395/22 and APP-12256/22). Edilberto O. Silva thanks colleagues at UFMA for valuable discussions. F. A. acknowledges the Inter University Center for Astronomy and Astrophysics (IUCAA), Pune, India, for granting a visiting associateship.

\bibliographystyle{apsrev4-2}
%

\end{document}